\documentclass[acmtog, authorversion]{acmartOURS}
\acmSubmissionID{171}  
\setcitestyle{square}

\acmJournal{TOG}

\setcopyright{acmcopyright}

\acmDOI{}

\acmISBN{XXXXXXXXXXXXX}

\acmConference[SIGGRAPH'18]{ACM SIGGRAPH conference}{July 2018}{Vancouver, Canada}

\usepackage[ruled]{algorithm2e} 
\usepackage{color}
\usepackage{steinmetz}
\usepackage{graphicx}
\usepackage{makecell}
\usepackage{amssymb}
\usepackage{amsmath}
\usepackage{mathtools}
\usepackage{multirow}
\usepackage{wrapfig}
\usepackage{algorithmic}
\usepackage{newlfont} 
\usepackage[english]{babel} 
\usepackage{enumitem}
\usepackage{microtype}
\usepackage{floatflt}
\usepackage{booktabs} 
\usepackage{subcaption}
\usepackage{epsfig}
\usepackage{xcolor}

\makeatletter
\newcommand\footnoteref[1]{\protected@xdef\@thefnmark{\ref{#1}}\@footnotemark}
\makeatother

\newcommand{\nothing}[1]{}

\newcommand{\rev}[1]{{ #1}}

\definecolor{YangColor}{rgb}{0.7,0,0} 

\definecolor{ZhanColor}{rgb}{0,0.6,0} 

\definecolor{VangelisColor}{rgb}{0,0,0.8} 

\definecolor{ChrisColor}{rgb}{0.9,0.44,0.0} 

\definecolor{KaranColor}{rgb}{0.56,0.34,0.62} 

\definecolor{SubhransuColor}{rgb}{1.0,0.3,0.3} 

\newcommand{\pseudocode}{Program}
\floatstyle{plain}
\newfloat{algorithm}{tbhp}{lop}
\floatname{algorithm}{\pseudocode}

\ifdefined\mainonly
\else
\newcommand{\mainonly}{0}
\fi

\makeatletter
\renewcommand{\paragraph}{%
  \@startsection{paragraph}{4}%
  {\z@}{1ex \@plus 1ex \@minus .2ex}{-1em}%
  {\normalfont\normalsize\it\indent}%
}
\makeatother

\newcommand{\bb}{\mathbf{b}}

\newcommand{\bm}{\mathbf{m}}

\newcommand{\bq}{\mathbf{q}}

\newcommand{\bv}{\mathbf{v}}

\newcommand{\bx}{\mathbf{x}}
\newcommand{\by}{\mathbf{y}}
\newcommand{\bz}{\mathbf{z}}

\newcommand{\bomega}{\mbox{\boldmath$\omega$}}

\newcommand{\bphi}{\mbox{\boldmath$\phi$}}

\newcommand{\btheta}{\mbox{\boldmath$\theta$}}
\newcommand{\bxi}{\mbox{\boldmath$\xi$}}

\makeatletter
\def\@copyrightspace{\relax}
\makeatother

\begin{document}

\title{VisemeNet:  Audio-Driven Animator-Centric Speech Animation}


\author{Yang Zhou}
\affiliation{%
 \institution{University of Massachusetts Amherst}
}
\email{yangzhou@cs.umass.edu}

\author{Zhan Xu}
\affiliation{%
 \institution{University of Massachusetts Amherst}
}
\email{zhanxu@cs.umass.edu}

\author{Chris Landreth}
\affiliation{%
 \institution{University of Toronto}
}
\email{c.landreth@rogers.com}

\author{Evangelos Kalogerakis}
\affiliation{%
 \institution{University of Massachusetts Amherst}
}
\email{kalo@cs.umass.edu}

\author{Subhransu Maji}
\affiliation{%
 \institution{University of Massachusetts Amherst}
}
\email{smaji@cs.umass.edu}

\author{Karan Singh}
\affiliation{%
 \institution{University of Toronto}
}
\email{karan@dgp.toronto.edu}



\begin{abstract}
We present a novel deep-learning based approach to producing animator-centric speech motion curves that drive a JALI or standard FACS-based production face-rig, directly from input audio.
Our three-stage Long Short-Term Memory (LSTM) network architecture is motivated by psycho-linguistic insights: segmenting speech audio into a stream of phonetic-groups is sufficient for viseme construction; speech styles like mumbling or shouting are strongly co-related to the motion of facial landmarks; and animator style is encoded in viseme motion curve profiles.  
Our contribution is an automatic real-time lip-synchronization from audio solution that integrates seamlessly into existing animation pipelines. 
We evaluate our results by: cross-validation to ground-truth data; animator critique and edits; visual comparison to recent deep-learning lip-synchronization solutions; and showing our approach to be resilient to diversity in speaker and language.

\end{abstract}

%
%

\ccsdesc[500]{Computing methodologies~Animation}
\ccsdesc[300]{Computing methodologies~Machine learning algorithms}

\keywords{facial animation, neural networks}

\begin{teaserfigure}
\centering
\vskip -3mm
\includegraphics[width=0.99\textwidth]{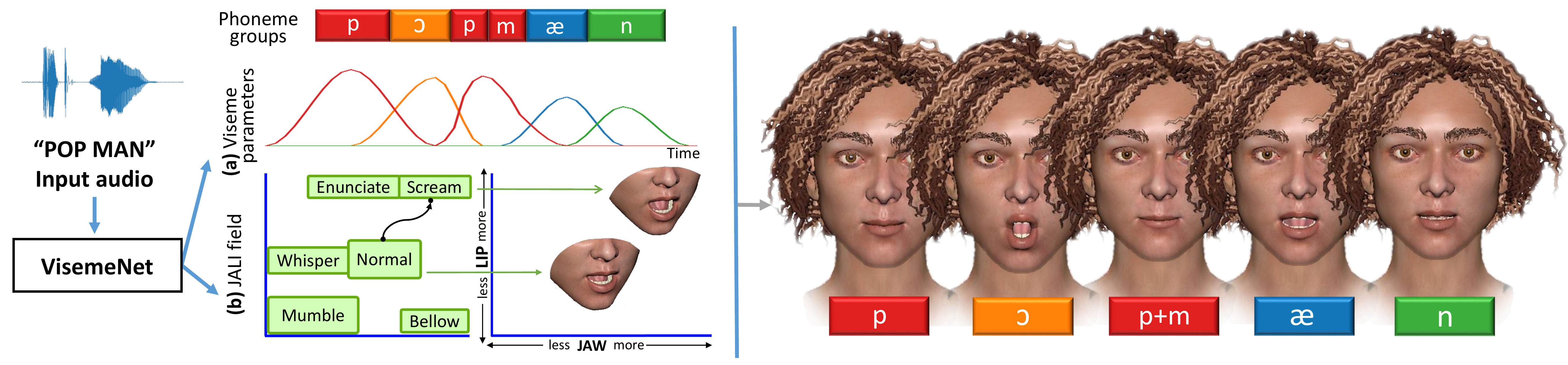}
\vskip -5mm
\caption{VisemeNet is a deep-learning approach that uses a 3-stage LSTM network, to predict compact animator-centric viseme curves with proper co-articulation, and speech style parameters, directly from speech audio in near real-time (120ms lag).}
\label{fig:teaser}
\end{teaserfigure}

\maketitle

\vskip -2mm
\section{Introduction}
The importance of realistic computer generated human faces cannot be understated. A diverse set of applications ranging from entertainment (movies and games), medicine (facial therapy and prosthetics) and education (language/speech training and cyber-assistants) are all empowered by the ability to realistically model, simulate and animate human faces. Imperfect emulation of even subtle facial nuance can plunge an animated character into the Uncanny Valley, where the audience loses trust and empathy with the character. Paradoxically, the greater the rendered realism of the character, the less tolerant we are of flaws in its animation \cite{MacDorman2009}. The expressive animation of speech, unsurprisingly, is a critical component of facial animation that has been the subject of research for almost half-a-century \cite{baillybook}.

Typically, keyframing or performance capture are used for high-end commercial animation. Keyframing by professional  animators is both expressive and editable, but laborious and prohibitively expensive in time and effort. Performance capture solutions are the opposite; recording a vocal performance is relatively easy, but hard to further edit or refine, and voice actors often have visually inexpressive faces.

Recent approaches however, have shown that there is enough detail in the audio signal itself to produce realistic speech animation. JALI \cite{Edwards:2016:JAV} presented a FACS \cite{Ekman:1978wx} and psycho-linguistically inspired face-rig capable of animating a range of speech styles, and an animator-centric procedural lip-synchronization solution using an audio performance and speech transcript as input. Following JALI, a number of conncurrent approaches have successfully shown the use of deep learning to capture a mapping between input audio and an animated face   \cite{Karras:2017:AFA,Suwajanakorn:2017:SOL,Taylor:2017:DLA}. These approaches however, produce results that, unlike JALI, do not integrate well into a typical animator workflow. This paper addresses the problem of producing animator-centric speech animation directly from input audio.

Arguably, existing deep learning based solutions can be used to first generate a text transcript \cite{graves2014towards} from audio input, and then align it with the audio \cite{mcauliffe2017montreal}, resulting in the input required by JALI. As shown in the evaluation (Section 6), dissecting the problem into  these three independent modules is error prone, and further any solution involving a speech transcript becomes strongly language and lexicon specific.
In contrast, Karras et al. \cite{Karras:2017:AFA} show that an audio signal has the potential to predict an animated face, agnostic of language.

Our approach is inspired by the following psycho-linguistic observations:\\
- While classification of a precise phonetic stream from audio can be difficult due to categorical perception \cite{fugate2013}, and its dependence on cultural and linguistic context, our problem of predicting a stream of phoneme-groups is simpler. Aurally individual phonemes within a phoneme-group are often hard to distinguish (eg. {\em pa} and {\em ba}) \cite{Liberman1957}, but unnecessary for speech animation as they map to near identical visemes \cite{fisher1968confusions}.\\
- The Jaw and Lip parameters in the JALI model that capture speech style as a combination of the contribution of the tongue-jaw and face-muscles to the produced speech, are visually manifested by the motion of facial landmarks on the nose, jaw and lips.\\
- The profile of speech motion curves (attributes like onset, apex, sustain and decay) capture speaker or animator style in professionally keyframed animation. Learning these curves from training data allow us to encode aspects of animator or speaker style.

We thus propose a three-stage network architecture: one that learns to predict a sequence of phoneme-groups from audio; another that learns to predict the geometric location of important facial landmarks from audio; and a final stage that learns to use phoneme-groups and facial landmarks to produce JA-LI parameter values and sparse speech motion curves, that animate the face.

The contribution of this paper is thus a deep-learning based architecture to produce state-of-the-art animator-centric speech animation directly from an audio signal in near real-time (120ms lag). Our evaluation is fivefold: we evaluate our results quantitatively by cross-validation to ground-truth data; we also provide a qualitative critique of our results by a professional animator; as well as the ability to further edit and refine the animated output; we also show our results to be comparable with recent non-animator-centric audio to lip-synchronization solutions; finally we show that with speech training data that is reasonably diverse in speaker, gender and language, our architecture can provide a truly language agnostic solution to speech animation from audio.

\vskip -2mm
\section{Related Work}
The large body of research on audiovisual speech animation \cite{baillybook} can be broadly classified into \textit{procedural},   \textit{performance-capture}, \textit{data-driven}, and more specifically the very recent \textit{deep-learning} based techniques.
We focus on techniques based on deep-learning, that are of greatest relevance to our research, after a brief overview of the other three approaches (refering the reader to \cite{Edwards:2016:JAV} for a more detailed review).

\paragraph{Procedural.}

\begin{figure}[t!]
\centering
\includegraphics[width=\linewidth]{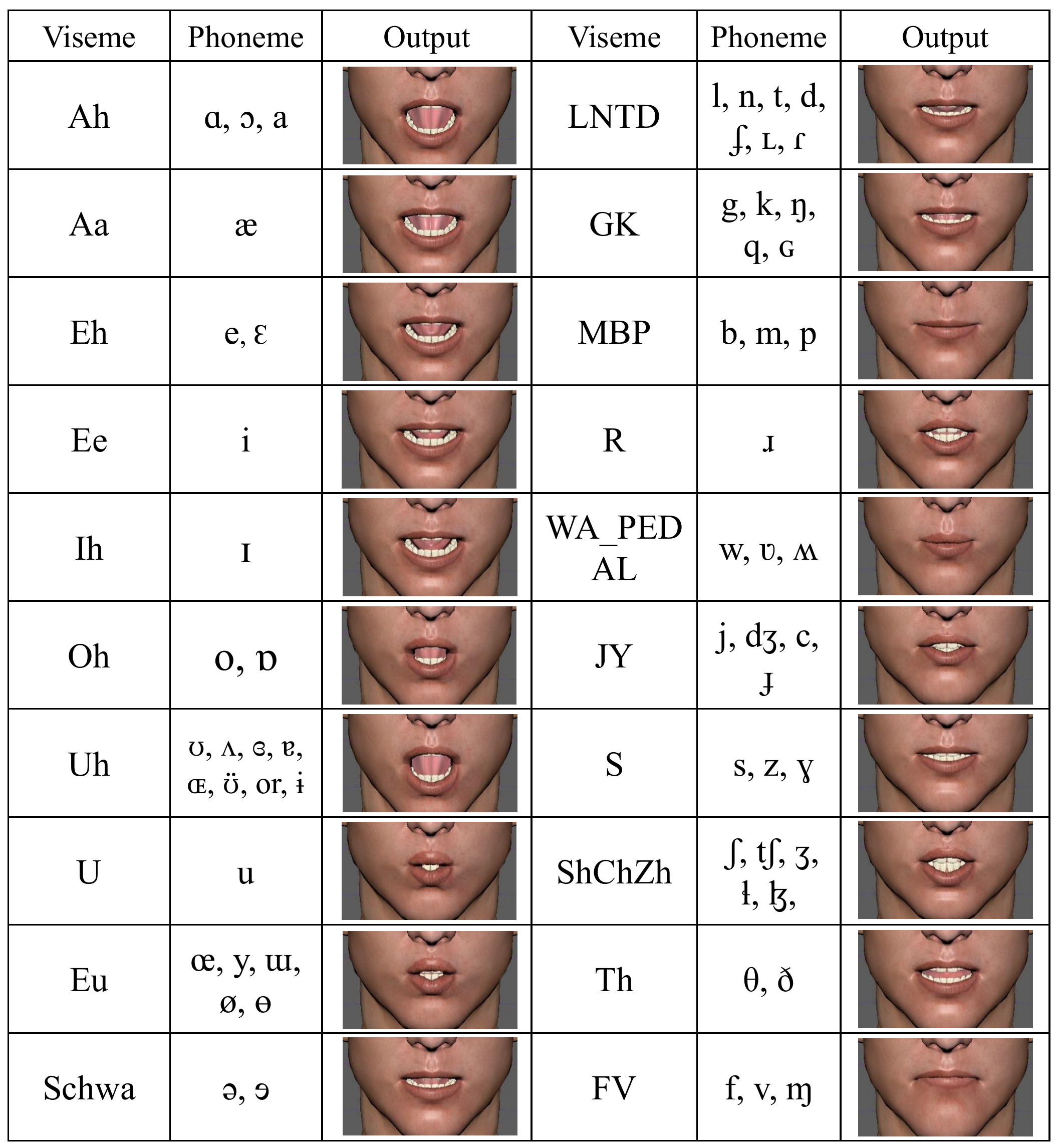}
\vskip -4mm
\caption{ List of visemes along with groups of phonemes (in International Phonetic Alphabet format) and corresponding lower face rig outputs that our architecture produces.}
\vskip -2mm
\label{fig:viseme_list}
\end{figure}

Procedural speech animation segments speech into a sequence of phonemes, which are then mapped by rules to visemes. A \textit{viseme} or \textit{visible phoneme} \cite{fisher1968confusions} refers to the shape of the mouth at the apex of a given phoneme (see Figure~\ref{fig:viseme_list}). The three problems that a procedural approach must solve are: \textit{mapping} a given phoneme to a viseme (in general a many-many mapping based on the spoken context \cite{Taylor:2012}); \textit{co-articulation}, or the overlap in time between successive visemes, resulting from the fluid and energy efficient nature of human speech, often addressed using Cohen and Massaro's \shortcite{Cohen:1993tw} seminal {\em dominance model}; \textit{viseme profile}, or the curve shape that defines the attack, apex, sustain and decay of a viseme over time \cite{Bailly:1997hv} . JALI \cite{Edwards:2016:JAV} defines the state of the art in procedural speech animation, producing compact animator-friendly motion curves that correspond directly to the input phonetic stream.

\paragraph{Performance-capture.}

Performance-capture based speech animation transfers motion data captured from a human performer onto a digital face \cite{Williams:1990}. 
Performance capture has become increasingly powerful and mainstream with the widespread adoption of cameras, depth sensors, and reconstruction techniques, that can produce a 3D face model from a single image \cite{singleimage:2017}. 
Real-time performance-based facial animation research \cite{Weise:2011jd,Li:2013jn} and products like {\em Faceware} (\textit{faceware.com}), are able to create high quality general facial animation, including speech, and can be further complemented by speech analysis \cite{Weise:2011jd}.
The disadvantage of performance capture is that is visually limited by the actor's performance and is difficult for an animator to edit or refine.

\paragraph{Data-driven.} 

These approaches smoothly stitch pieces of facial animation data from a large corpus, to match an input speech track \cite{Bregler:1997}, using morphable \cite{Ezzat:2002}, hidden Markov \cite{Wang:2012uq}, and active appearance models (\textsc{aam}) \cite{Anderson:2013wv,Taylor:2012}. These data-driven methods tend to be limited in scope to the data available, and the output, like performance-capture, is not animator-centric.

\paragraph{Deep learning-based speech animation.}

Recent research has shown the potential of deep learning to provide a compelling solution to automatic lip-synchronization simply using an audio signal with a text transcript \cite{Taylor:2017:DLA}, or even without it \cite{Karras:2017:AFA}.

Taylor et al.'s approach \shortcite{Taylor:2017:DLA} requires an input text transcript, which even if automated \cite{graves2014towards}, introduces more complexity and language dependence than we believe is necessary for animating speech. 
The predicted output is an 8-dimensional face-space, that is then re-targeted to any animation rig. While the rig can be post-edited by an animator, there is no compact  mapping from audio to animator-centric viseme curves or facial action units. As a result while it is possible to overlay the speech output with expression, it is harder to edit or refine the animation of the spoken content or its style.
\rev{In contrast, our network is trained to map audio to
viseme animation curves, which are both sparse (i.e., very few viseme controls are active at a time, have non-zero value only for short periods), and are low-dimensional (the number of visemes is small). Our viseme curves are  more directly related to speech, compared to the viseme-agnostic mouth pose parameterization used in \cite{Taylor:2017:DLA}.} 

Karras et al.'s approach  \shortcite{Karras:2017:AFA} removes any dependence on a phonetic transcript, but is designed to directly output the vertex positions of a 3D mesh. While they produce impressive results from a small training set, their output is not well suited to current animation practice. Arguably, vertex positions or facial landmarks could be re-targeted to any facial animation rig \cite{Ribera2017}, yet the re-targeted result is not animator-centric. \rev{In addition, breaking the rig prediction into  separate steps is problematic for a number of reasons. First, face rigs have lots of redundancy, thus a per frame best fit to rig space can have temporal discontinuities. Second,  rig animation curves are often correlated. Independent editing or sparsification of these curves can break perceptual constraints (e.g., a closed mouth on bilabial sounds). Third, accumulated errors in vertex positions  can be exacerbated after subsequent fitting steps. In contrast, our method does not break rig prediction into separate, independently optimized steps. Our  whole network with all its stages  is  trained as a single  pipeline    optimized to minimize animation error.}

Suwajanakorn et al. \shortcite{Suwajanakorn:2017:SOL} present a deep network to synthesize an output video of Barack Obama, combining an input audio and target video of him speaking. While their problem statement is quite different from ours, we are encouraged by their successful use of a Long-Short Term Memory (LSTM) network for lip-synchronization, and adopt it for the stages of our overall network.

\rev{From a technical point of view, our proposed network architecture differs from  the above deep learning approaches in several  key design aspects. It  employs a three-stage  network architecture to (a) segment speech audio into streams of phonetic groups related to viseme construction, (b) predict facial landmarks capturing speech style cues, (c) then uses the extracted phoneme groups, landmarks and audio features to produce viseme motion curves. All  stages are necessary to achieve high performance, as discussed in our experiments section (Section \ref{sec:results}). Furthermore, we combine different sources of audiovisual data for training. Our training data include  audio clips with ground-truth phonemes,  video clips with tracked landmarks, and 3D facial animations with ground-truth visemes. All stages are jointly trained using  multi-task learning to minimize errors  in the prediction of phoneme groups, landmarks, visemes, co-articulation, and  speech style parameters.  }

\begin{figure*}[htbp]
  \centering
  \includegraphics[width=\textwidth]{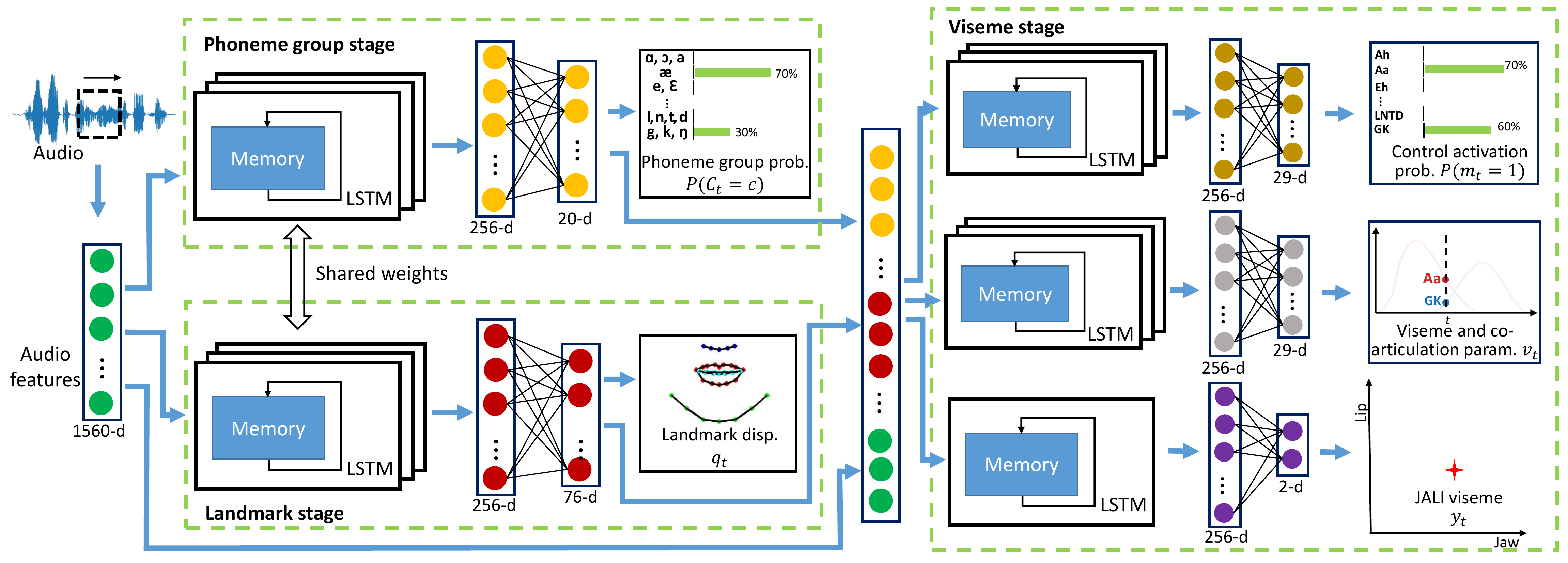}
  \vskip -4mm
  \caption{Our architecture processes an audio signal (left) to predict JALI-based viseme representations: viseme and co-articulation control activations (top right), viseme and co-articulation rig parameters (middle right), and   2D\ JALI viseme field parameters (bottom right). Viseme prediction is performed in three LSTM-based stages: the input audio is first processed through the phoneme group  stage (top left) and landmark stage (bottom left), then the predicted phoneme group and landmark representations along with  audio features are processed through the viseme  stage.\  }
  \vskip -1mm
  \label{fig:architecture}
\end{figure*}

\vskip -2mm
\section{Algorithm Design}
\label{sec:overview}

Our approach is designed to achieve high-quality, animator-editable, style-aware, language agnostic, real-time 
speech animation from audio.

Following the current animation practice of FACS-like face rigs, and state-of-the-art animator-centric speech \cite{Edwards:2016:JAV}, our network uses an input audio signal to predict a sparse and compact set of viseme values (see Figure~\ref{fig:viseme_list}), and jaw and lip parameters, over time. Our neural network  architecture (see Figure \ref{fig:architecture}) is designed to exploit psycho-linguistic insights and make the most effective use of our training data.

\paragraph{Phoneme group prediction.} A large part of our network, the ``phoneme group stage'' in Figure \ref{fig:architecture} (top left box),  is dedicated to map audio to phonemes groups corresponding to  visemes. For example, the two labio-dental  phonemes  $/f$ and $v/$ form a group that maps to a single, near-identical viseme \cite{Edwards:2016:JAV}, where the lower lip is pressed against the upper teeth in Figure \ref{fig:viseme_list} (last row, right). We identified $20$ such visual  groups of phonemes expressed in  the International Phonetic Alphabet (IPA)\ in Figure \ref{fig:viseme_list}. Our network is trained to recognize these phoneme groups from audio without  any text or phonetic transcript. By predicting only phonetic groups relevant to animation, our task is simpler and less sensitive to linguistic context. The network can also be trained with less data than most speech processing pipelines. As demonstrated in the evaluation (Section 6), using off-the-shelf audio-to-text techniques to extract individual phonemes that subsequently predict visemes, leads to  lower performance than our architecture.

\paragraph{Speech style prediction.}  Phoneme  groups alone have no information of vocal delivery and cannot predict visemes, specially for expressive emotional speech. The same phoneme /\ae/ (see Figure \ref{fig:teaser}b) might be pronounced conversationally, or screamed. These style attributes of the audio performance  can be captured using jaw and lip parameters \cite{Edwards:2016:JAV}. 
These parameters are also strongly correlated to the (2D frontal) jaw and lip landmarks in a visual capture of the vocal performance. The ``landmark  stage'' of our network in Figure \ref{fig:architecture} (bottom-left box) is designed to predict a set of jaw and lip landmark positions over time given input audio.

\paragraph{Viseme prediction.} The last part of our network, the ``viseme stage'' in Figure \ref{fig:architecture}(right-box), combines the intermediate predictions of phoneme groups, jaw and lip parameters, as well as the audio signal itself to produce visemes.
By training our architecture on a combination of  data sources containing \ audio, 2D\ video, and 3D\ animation of human speech, we are able  to predict visemes accurately. We represent visemes based on the JALI  model \cite{Edwards:2016:JAV}, comprising a set of intensity values for $20$ visemes and $9$ co-articulation rules, and \textbf{JA}W and \textbf{LI}P parameters that capture speaking styles. The viseme and co-articulation values over time can animate standard FACS-based production rigs, with the JA-LI parameters for a rig adding control over speech style. Notably, these are precisely the controls professional animators keyframe, and are thus directly amenable to editing and refinement.

\paragraph{Transfer learning from audiovisual datasets.} 

We need reliable sources of diverse training data with compelling variation in terms of different speakers, speech styles, and emotional content, to train a network that generalizes well.
One potential source of training data would be audio clips with corresponding streams of 3D facial rig parameters. 
Such a large coherent dataset with enough variability is not easily available, and would be too expensive to create with 
professional animators.  
On the other hand, there is a wealth of publicly available audiovisual corpora (2D audio+video+text transcript clips), such as BIWI   \cite{eth_biwi_00760}, SAVEE \cite{HaqJackson_AVSP09}, and GRID \cite{cooke2006audio}. 
Although these corpora do not contain any facial rig parameters, they are nonetheless valuable, in the context of our network architecture:\\
- The faces in the video clips can be accurately detected and annotated with landmarks through modern computer vision. The extracted 
facial landmarks  corresponding to the speech audio are then useful to train the landmark stage of our network.\\
- The text transcripts can be automatically aligned with the audio clip \cite{mcauliffe2017montreal}, to provide training phonemes  for the phoneme group stage of our network.\\ 

To make use of the large amounts of data in these audiovisual datasets, we employ a \emph{transfer learning} procedure to train our network. We first ``pre-train'' the phoneme group and landmark stages of our network based on training phoneme groups and landmarks extracted from the above audiovisual datasets. 
We then initialize  these two stages according to their pre-trained parameters, and then jointly train the whole network. To perform this joint training, we still need a dataset of audio clips with associated streams of facial rig parameters. 
Mapping phoneme groups and facial landmarks to visemes however, is significantly easier than mapping general speech audio to phonemes and landmarks. Further, the phoneme groups are strongly correlated to visemes, while the landmarks are strongly correlated to jaw and lip parameters. Thus, to train the viseme stage, a much smaller dataset of example 3D animations with face rig parameters is required for sufficient generalization. We empirically observed that using our transfer learning procedure results in a better generalization performance than simply training the entire network  on a  small dataset of audio clips with rig  parameters. We also found that adapting a Multi-Task Learning (MTL) procedure to train the network simultaneously according to multiple objectives involving phoneme group, landmark, viseme and  other rig parameter prediction
was also important to achieve high performance. 

\paragraph{Memory-enabled networks.} We adapt a memory-enabled neural network architecture based on Long Short-Term Memory units (LSTMs)  \cite{lstm_tutorual,hochreiter1997long}
for all  stages of our network. We believe that memory-based networks are important to correctly capture co-articulation and speech context from input audio, which even for a human listener, is challenging from isolated audio fragments. Memory-enabled networks explicitly store and represent a large amount of context in the input signal that is useful to reliably infer the spoken context. Finally, another advantage of our architecture is that it can predict viseme motion curves in near real-time (120ms or 12 frame lag), given the input audio on modern GPUs.
In the following sections, we discuss the network architecture and training procedure in more detail.

\vskip -2mm
\section{Network architecture}
\label{sec:technical}

Our network architecture  takes an audio signal as input and outputs viseme representations based on JALI. As discussed in the previous section and shown in Figure \ref{fig:architecture}, our network has a three stage-architecture: the input audio is first processed through the phoneme group  and landmark stages, then the predicted phoneme group and landmark representations along with  audio features are processed through the viseme prediction stage. All branches are based on a combination of memory-based LSTM units, which encode context
in the input signal, and fully connected layers, which decode the  memory of the units into  time-varying predictions. Below we discuss our input audio representation and the three stages of our network in more detail.  

\paragraph{Input audio representation.} Given an audio signal as input, we extract a feature vector for each frame encoding various power spectrum and  audio frequency signal characteristics. Our feature vector concatenates $13$  Mel Frequency Cepstral Coefficients (MFCCs) \cite{Davis80comparisonof} that have been widely used for speech recognition, $26$ raw Mel Filter Bank (MFB) features that have been shown to be particularly useful for emotion discrimination in audio \cite{Busso2007}, and finally $26$ Spectral Subband Centroid features that often result in better speech recognition accuracy when they are used in conjuction with MFCCs  \cite{paliwal1998spectral}. The resulting 65-dimensional feature vector is passed as input to all three stages of our network for further processing. The features are extracted every $10$ ms, or in other words feature extraction is performed at a $100$ FPS rate. The  frequency analysis is performed within windows of size $25$ ms in the input audio. 

\paragraph{Phoneme group stage.} The phoneme group stage takes as input the audio features concatenated from $12$  frames before the current frame, and also the audio features of the current frame plus $11$ frames after the current one. This means that given real-time audio inputs, the network will infer visemes with a  lag of $120$ ms, plus the required time to extract audio features and perform viseme inference given  the audio features per frame (feature extraction and network inference take $1$ ms per frame  measured on a TitanX GPU, allowing real-time inference with the abovementioned lag). 
 The concatenation produces a $1560$-dimensional
sharfeature vector $\bx_t$ per frame $t$ (  $65$ features x $24$ frames)   covering audio signal information in a window of $240ms$. The rationale behind using this window is that it approximately matches the average duration of a phoneme in normal speech. We also found that that such window size represented a good trade-off between fast processing time and  high phoneme group prediction accuracy. 

The feature vector  $\bx_t$  passes  through three layers of unidirectional LSTM units that hierarchically  update their internal memory state (encoded with a 256-dimensional vector in our implementation) based on the input  information in the audio features. We found that  at least three layers (i.e., a deep network) were necessary to achieve sufficient generalization. The LSTM\ units can choose to either store  in their memory cells  representations of the incoming features, or alternatively erase representations from their memory cells. The choices of erasing or storing, as well as the transformations of the input features are controlled through non-linear functions (sigmoid and hyperbolic tangent functions)\ with learnable parameters (for their exact form, we refer  to the popular tutorial \cite{lstm_tutorual} and \cite{hochreiter1997long}). 

At each frame, the memory state of the last LSTM layer is decoded  towards probabilistic predictions of phoneme groups. The decoding is performed through two non-linear transformations. The first transformation involves a fully connected layer that takes as input the representation of the uppermost LSTM layer, applies a linear transformation on it to produce a $256$-dimensional  output, which is further processed through the commonly used REctified Linear  Unit (RELU) non-linearity. The second layer takes the resulting vector, applies another linear transformation on it producing a $20$-dimensional vector $\bz_t$, which is then passed through a softmax function to output a per-frame probability for each phoneme group listed in Figure \ref{fig:viseme_list}.

Overall, this network stage can be seen as a non-linear function $f$ that considers   audio features up to the current frame $\bx_{1:t}$ (by exploiting the  LSTM recurrent connections) to output phoneme group probabilities: $P(C_t = c)=f( \bx_{1:t},\btheta,\bphi)$ where  $C_t$ is a discrete random variable whose possible values $c$ are our phoneme groups,  $\btheta$ are the LSTM parameters, and $\bphi$ are the decoder parameters.
\rev{We note that we also experimented with a Bidirectional LSTM, as proposed in \cite{graves2005framewise}, yet we did not perceive any noticeable differences in the output predictions. We suspect this is because we consider audio features from a large window containing both past and future frames.}

\begin{figure}[t!]
\centering
\includegraphics[width=0.9\linewidth]{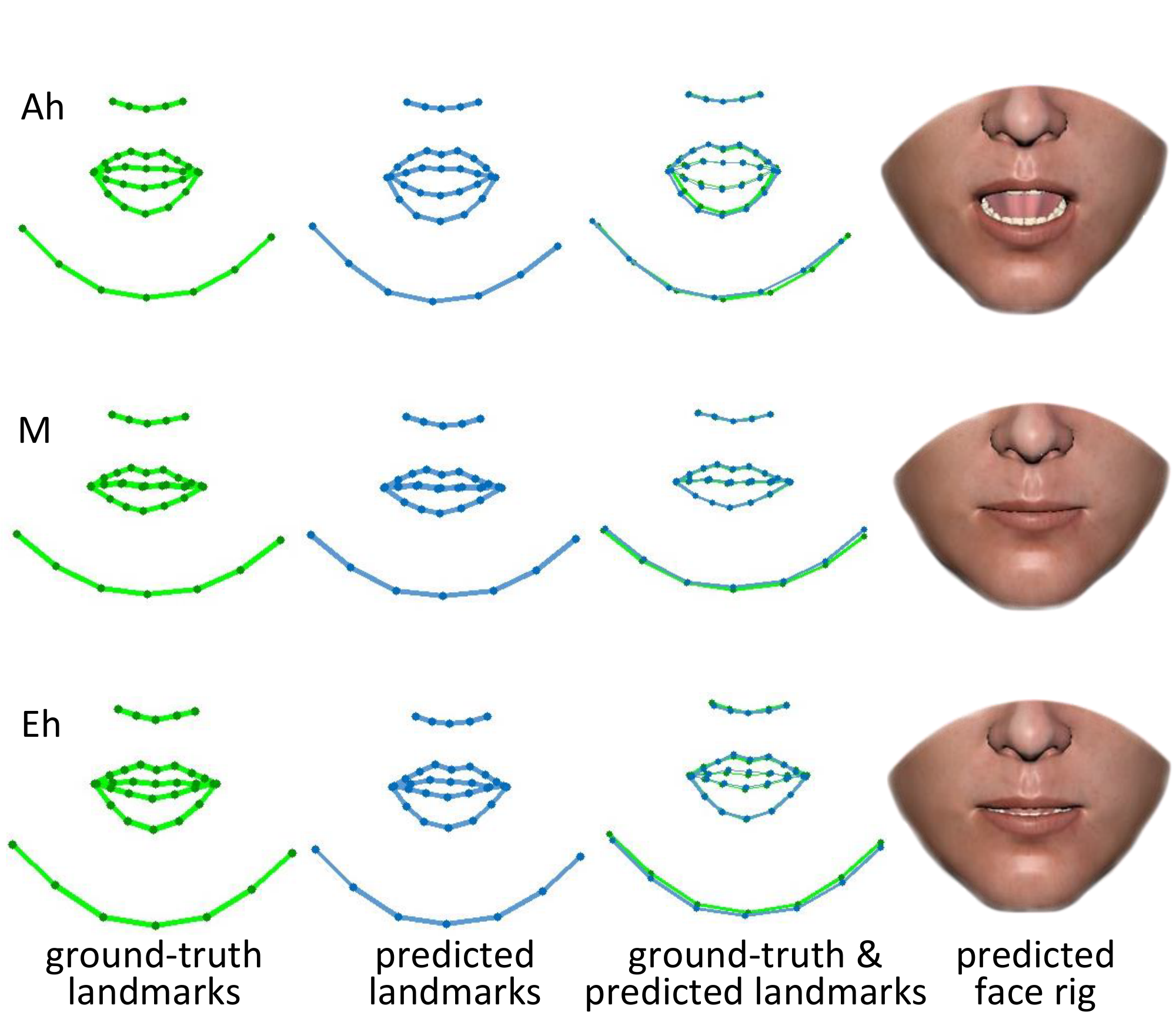}
\vskip -3mm
\caption{Landmark output predictions for a few test frames. The spoken phoneme is shown on the top left. The first column shows ground-truth landmarks. The second column shows landmark predictions from the landmark stage of our network for the same frames. \rev{In the third column the predictions are overlaid on the ground-truth to show differences. The last} column shows the corresponding predicted rig outputs. }
\vskip -2mm
\label{fig:landmarks}
\end{figure}

 \paragraph{Landmark stage.} This stage of our network takes as input the $1560$-dimensional feature vector $\bx_{t}$  per frame (same input as in the phoneme group part), passes it through three LSTM layers and decodes the uppermost layer memory  into  a sparse set of $38$ 2D\ facial landmarks, representing the jaw, lip, and nose configuration per frame. Ground-truth and predicted landmarks are visualized in Figure \ref{fig:landmarks}. The landmarks do not aim at capturing the morphology of a particular face, but instead approximately capture the shape of the lips, positions of jaw and nose of an average face. We found that predicting these visual cues are particularly useful to infer correct visemes that reflect speech style (e.g., mumbling, screaming). In particular,  the advantage of using these visual cues is that we can exploit  audio and  landmarks extracted from video available in large, public audiovisual datasets as additional supervisory signal to train the LSTM layers, as described in the next section. 

Since phonetic groups and lower face shape are correlated, the three LSTM layers are shared between the phoneme group and landmark stages. Sharing representations is a common strategy in multi-task learning \cite{Caruana:1997:ML}, which helps generalization when tasks are correlated. The decoder of the landmark stage is specific to landmark predictions and has its own learned parameters. It is composed of two  transformations, implemented as a fully connected layer followed by RELUs, and a second fully connected layer that outputs landmark displacements per frame. The displacements are expressed relative to landmarks representing an  average lower face in neutral expression. The displacements are stored in a $76$-dimensional vector $\bq_t$ per frame $t$, which simply concatenates displacement coordinates of all the landmarks. Given the neutral face landmarks $\bb$, the animated landmark positions can be computed as $\bb+\bq_t$ per frame. Overall, this part of our network can be seen as another non-linear function $h$ that considers  audio features up to the current frame $\bx_{1:t}$  and outputs landmark displacements per frame: $\bq_t = h( \bx_{1:t},\btheta, \bomega)$ where $\btheta$ are the shared LSTM parameters, and $\bomega$ are the decoder parameters of this stage.

\paragraph{Viseme  stage.} The viseme stage takes as input the produced phoneme group representations  $\bz_t$ (i.e., phoneme group activations before applying softmax),   landmark displacements $\bq_t$, and also the audio features $\bx_{t}$ and outputs JALI-based rig parameters and controls that determine the visemes per \ frame. Here, we use the audio features as additional input since phoneme groups and landmarks might not entirely capture all speech style-related information existing in audio (for example, fast breathing due to a nervous style of speech will not be captured in landmarks or phonemes, yet will manifest in audio features). 

The viseme stage produces the following outputs per frame $t$ (see also Figure \ref{fig:architecture}, right):

 (a) $29$  continuously-valued viseme animation and co-articulation parameters present in JALI (we refer to \cite{Edwards:2016:JAV} for more details). The rig parameters are represented by a $29$-dimensional continuous vector $\bv_t$.

(b) $29$ binary random variables, represented as a vector $\bm_t$, that indicate whether each of the above viseme and co-articulation control parameters is active per frame. The underlying reason for using these binary variables is that the activations of viseme and co-articulation rig parameters are largely sparse (Figure \ref{fig:teaser}a), since at a given frame only one viseme is dominantly active. If we train the network to match the viseme and co-articulation parameters in the training data,
without considering these binary indicator variables, then the predicted values of these action units are often biased towards  low or zero values since most of the time their corresponding training values
are zero. 

(c) an output $2$-dimensional vector  $\by_t$ representing the 2D JALI viseme field values capturing speech style per frame $t$.

Given the inputs  $\{ \bz_t, \bq_t, \bx_{t}\}$ concatenated as a single vector, the viseme stage uses a three-layer LSTM-based architecture
to produce the above outputs. Here, we found that using separate LSTM\ layers  (i.e., without shared parameters)\  for each output type offers the best performance, probably due to the fact that the JALI viseme field is  designed to be  independently controllable from the rest of rig parameters \cite{Edwards:2016:JAV}, and also because the continuous rig parameter values vary widely given their  activation state. Each LSTM\ layer uses\ units with a $256$-dimensional memory state. The rig parameters  $\bv_t$ and JALI viseme field values $\by_t$  are computed by decoding their corresponding uppermost LSTM layer memory through two dedicated fully connected layers with a RELU non-linearity in-between.  The binary  indicator variables are predicted by decoding their corresponding uppermost LSTM layer memory, followed by two dedicated fully connected layers with a RELU non-linearity in-between, and finally a sigmoid function that produces the probability of each rig parameter to be active or not. At test time, we first predict these activation probabilities. Then we produce the continuous values of the corresponding viseme and co-articulation rig parameters
whose activation probability is above a threshold, which we automatically set  during training.

Overall, this part of our network can be seen as a set of three non-linear functions $g_1,g_2,g_3$ that considers phoneme group predictions $\bz_{1:t}$, landmark predictions $\bq_{1:t}$, and audio features $\bx_{1:t}$ up to the current frame and output probabilities of rig parameter activations $P(\bm_t)= g_1(\bz_{1:t},\bq_{1:t},\bx_{1:t},\bxi_1)$, rig parameter values $\bv_t= g_2(\bz_{1:t},\bq_{1:t},\bx_{1:t},\bxi_2)$, and viseme values  $\by_t= g_3(\bz_{1:t},\bq_{1:t},\bx_{1:t},\bxi_3)$, where $\bxi=\{\bxi_1,\bxi_2,\bxi_3\}$ are learned parameters of each corresponding set of LSTM layer and decoder.
In the following section, we describe how all the parameters of our network are learned.

\vskip -2mm
\section{Training}
\begin{figure}[t!]
\centering
\includegraphics[width=1.0\linewidth]{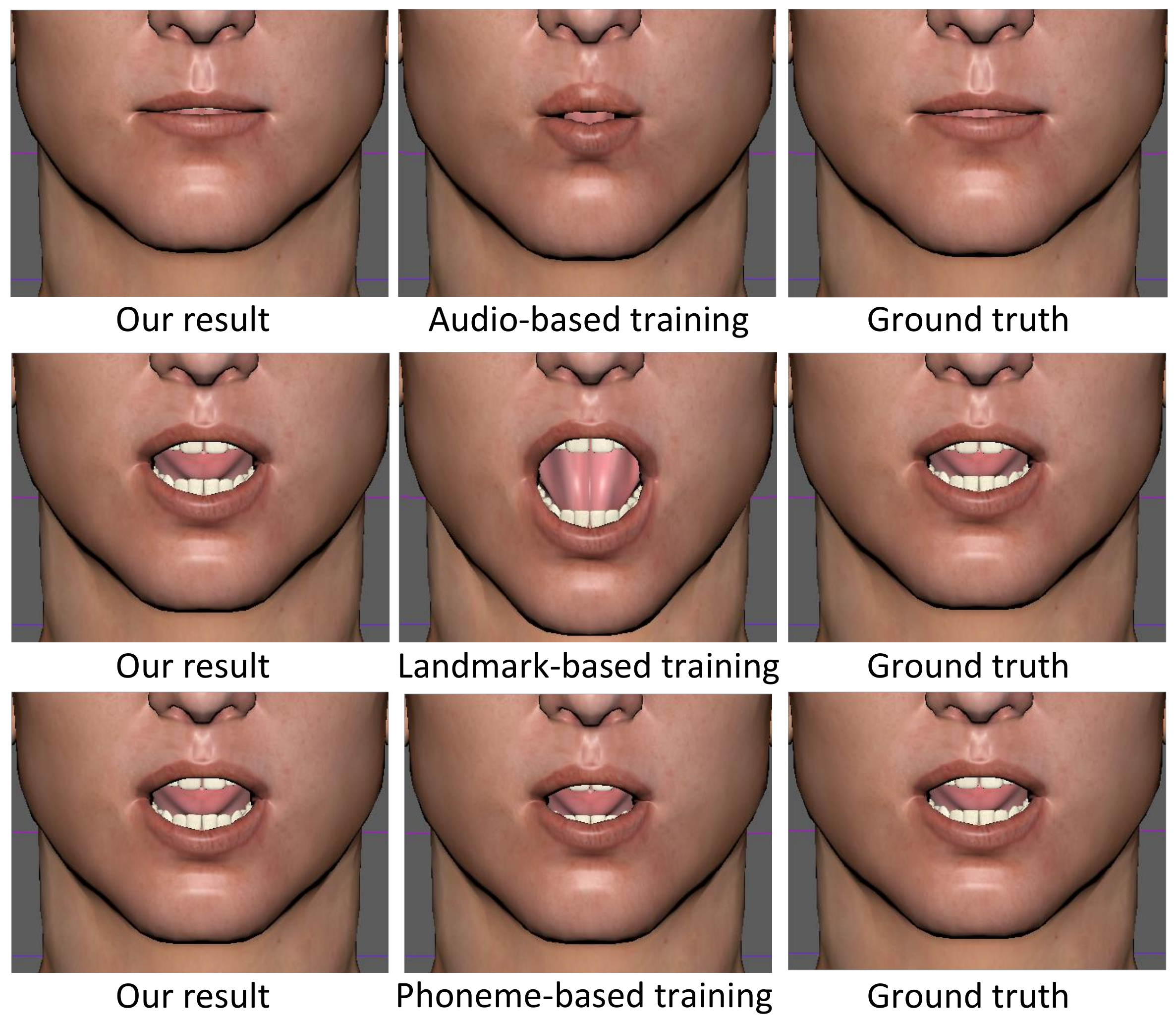}
\vskip -3mm
\caption{Characteristic outputs of alternative architectures versus our method for a frame from our test splits. \textbf{(top)} Comparison with using audio features alone (``audio-based'' training) \textbf{(middle)} Comparison with using landmarks and audio features alone (``landmark-based'' training) \textbf{(bottom)} Comparison with using phoneme groups and audio features alone (``phoneme-based'' training). Ground-truth for the corresponding test frame is shown on the right column. We refer to the video for the full clip.
}
\vskip -3mm
\label{fig:alternative_results_failures}
\end{figure}

We follow a two-step, transfer learning procedure to train our network.
Motivated by the observation that there are large valuable  amounts of audiovisual data with text transcripts available in public repositories, we first ``pre-train'' the phoneme group and landmark stages of our network based on $15$ hours of recorded audio along with ground-truth phoneme groups and tracked facial landmarks from video as supervisory signal. After this pre-training step, we fine-tune the whole network jointly to accurately predict visemes based on a smaller, painstakingly created dataset
consisting of one hour of audio with exemplar streams of JALI\ rig parameter values. Below we explain our datasets and pre-processing, then we discuss pre-training and joint training procedures. Training and test splits are discussed in the results section.

\paragraph{Audiovisual dataset.} This dataset contains audiovisual data from three repositories. First, we use the  GRID dataset \cite{cooke2006audio}, which contains transcripts, audio and video recordings of $1000$ sentences spoken by $34$ speakers ($18$ male, $16$ females) in a neutral style of speech with total time duration of about $14$ hours. The sentences
are  deliberately chosen to cover common phonemes in English. Second, we use the SAVEE dataset \cite{HaqJackson_MachineAudition10}, which contains transcripts, audio and video recordings of $480$ sentences spoken by $4$ male speakers expressing different emotions, including anger, disgust, fear, sadness and surprise.
The total clip duration is $30$ minutes.
Third, we used the BIWI 3D audiovisual corpus dataset \cite{eth_biwi_00760}, which contains
transcripts, audio, video
and RGBD
recordings of  $1109$ sentences spoken by $14$  speakers (6 males and 8 females) in various emotional and neutral styles of speech with total time duration of about $1$ hour. We pre-processed the videos of these datasets to extract facial landmarks involving lips, jaw and nose (Figure \ref{fig:landmarks}, left column) through DLib \cite{dlib09} and FaceWare Analyzer \cite{faceware}. The  landmarks were aligned with our average face template in neutral poses and normalized to be invariant to face location, orientation and size. Then we extracted training landmark displacements for each frame relative to the average face. Given the provided text transcripts in these datasets, we  used the Montreal Forced Aligner (MFA) \cite{mcauliffe2017montreal} to align audio with text and extract phonemes along with corresponding phoneme groups.  

\paragraph{JALI-annotated dataset.} An experienced animator
created rigs according to the JALI\ template for the  BIWI  dataset (total $1h$ of rig motion curves with corresponding audio involving the  $14$ BIWI\ speakers).

\paragraph{Pre-training.} Given $N$ audio clips with sequences of landmark displacements $\hat{\bq}_{1:t_n}$ (where $t_n$ is number of frames of the $n^{th}$ audio clip, $n=1...N$), and corresponding phoneme groups $\hat{c}_{1:t_n}$  extracted per frame from the training audiovisual datasets, the goal of the pre-training step is to estimate the decoder parameters  $\bphi$ of the phoneme group stage, the decoder parameters  $\bomega$ of the  landmark stage, and the parameters
 $\btheta$ of their shared LSTM\ layers. The parameters are learned such that the predicted phoneme groups match the training ones as much as possible, the predicted landmark coordinates are as close as possible to the training ones, and also the predicted landmarks do not change over time abruptly. The goals can be expressed with a combination of a classification loss  $L_c(\btheta, \bphi)$  for phoneme groups, a regression loss  $L_q(\btheta, \bomega )$ for landmarks, and  another loss $L_q'(\btheta, \bomega )$ that promotes smoothness in the predicted landmark movement.
This combination is expressed as the following multi-task loss:
\begin{equation}
L_{1}(\btheta, \bphi, \bomega) = w_c\ L_c(\btheta, \bphi)+ w_q L_q(\btheta, \bomega)+ w_q' L_q'(\btheta, \bomega)
\label{eqn:pretrain_loss}
\end{equation}
where weights of the three losses are set as $w_c=0.75$, $w_q=0.25$, $w_q'=0.1$ in all our experiments, computed via hold-out validation. The classification loss $L_c(\btheta, \bphi)$ favors parameters that maximize the probability of the training phoneme groups, or equivalently minimize their negative log-probability. It can be expressed as the popular cross-entropy multi-class loss:
\begin{equation}
L_c(\btheta, \bphi) =- \frac{1}{N}\sum_{n=1}^N \bigg( \frac{1}{t_n}\sum_{t_{}=1}^{t_n} \log P(C_{t} = \hat{c}_{t}) \bigg)
\label{eqn:phoneme_loss}
\end{equation}
The regression loss $L_q(\btheta, \bomega )$ is expressed as the absolute differences (i.e., $L_1$-norm loss) between the training and predicted landmark coordinates (their total number is $M=76$ coordinates from $38$ 2D\ landmarks):
\begin{equation}
L_q(\btheta, \bomega) = \frac{1}{N}\frac{1}{M}\sum_{n=1}^N \bigg( \frac{1}{t_n}\sum_{t_{}=1}^{t_n} ||\bq_{t} - \hat{\bq}_{t}||_1 \bigg)
\label{eqn:landmark_loss}
\end{equation}
The smoothness loss $L_q'(\btheta, \bomega )$ penalizes large absolute values of  landmark motion derivatives with respect to time:
\begin{equation}
L_q'(\btheta, \bomega) = \frac{1}{N} \frac{1}{M}\sum_{n=1}^N \bigg( \frac{1}{t_n}\sum_{t=1}^{t_n}||\dot \bq_{t} ||_1 \bigg)
\label{eqn:landmark_smooth_loss}
\end{equation}
where  $\dot \bq_{t}$ represents the derivative of the predicted landmark displacements over time. The derivative  is computed through central finite differences in our implementation.

Minimizing the above multi-task loss function is done through batch gradient descent with batch size $256$, learning rate $0.00001$, momentum set to $0.9$, over $2M$ iterations.

\paragraph{Joint training.} For joint training, we initialize the parameters $\btheta,\bphi,\bomega$ of the phoneme group and landmark stages to their pretrained values, which
are already expected to be close to a desired local minimum. Then we estimate all the parameters of the whole network jointly,  including the parameters  $\bxi=\{\bxi_1,\bxi_2,\bxi_3\}$ of the viseme prediction branch, such that based on the JALI-annotated dataset, we satisfy the above goals:\ (a) the predicted  viseme and co-articulation parameter activations match the ground-truth ones through a binary classifications loss $L_a(\bxi_1),$ (b) the  predicted viseme and co-articulation parameters are as close as possible to the ground-truth ones when these units are active through a regression loss $L_v(\bxi_2)$ modified to consider these activations, (c) the predicted 2D\ JALI\ viseme field values are also as close as possible to the ground-truth ones through a regression loss $L_j(\bxi_3)$, (d) the rig parameters and JALI field values do not change abruptly over time through two smoothness losses $L_v'(\bxi_2)$ and $L_j'(\bxi_3)$, and finally (e) the predicted phoneme groups and landmarks remain close to the ground-truth ones (as done in the pre-training step). This goal can be expressed again as a multi-task loss: 
\begin{align}
L_{2}(\btheta, \bphi, \bomega, \bxi) &= L_{1}(\btheta, \bphi, \bomega) + w_a L_a(\bxi_1)+ w_v L_v(\bxi_2) \nonumber \\
& + w_j L_j(\bxi_3)  + w_v' L_v'(\bxi_2)+ w_j' L_j'(\bxi_3)  
\label{eqn:mtl_loss}
\end{align}
where $L_1(\btheta, \bphi, \bomega)$ is the loss of Eq. \ref{eqn:pretrain_loss} (same as pre-training, but now evaluated in the JALI-annotated dataset). The loss weights are set in all our experiments via hold-out validation as follows: $w_{a}=0.1, w_v=0.2, w_j=0.2, w_v'=0.15$, and $w_j'=0.15$. Note that this loss function is not decomposable because the predictions (and in turn, the losses) associated with the viseme  branch depend on the predicted phonemes and landmarks of the other two stages during training. Below we describe the individual loss functions in detail.

The  loss function $L_a(\bxi_1)$ penalizes disagreements between  predicted parameter activations $\bm_{t}$  and ground-truth parameter activations $\hat{\bm}_{t}$ for each training frame $t$. Since multiple rig parameters can be active at a given time, this loss function attempts to maximize the probability of correct, individual activations per parameter (or equivalently minimize their negative log-probability). It can be expressed as a sum of $A=29$ binary cross-entropy losses, one per rig parameter: 

\begin{align}
L_a(\bxi_1) =- \frac{1}{N}\frac{1}{A}\sum_{n=1}^N \sum_{a=1}^N \bigg( \frac{1}{t_n} & \sum_{t_{}=1}^{t_n} [\hat{m}_{a,t}=1] \log P(m_{a,t} = 1) \nonumber  \\
\frac{1}{t_n} & \sum_{t=1}^{t_n} [\hat{m}_{a,t}=0] \log P(m_{a,t} = 0)
\bigg)
\label{eqn:activation_loss}
\end{align}
where $[\hat{m}_{a,t}=1], [\hat{m}_{a,t}=0]$ are binary functions indicating whether the rig parameter $a$ is active or not at frame $t$. 

The loss function $L_v(\bxi_2)$ measures absolute differences between the training values $\hat{v}_{a,t}$ and predicted values $v_{a,t}$ of each viseme and co-articulation rig parameter $a$ when these are active according to the ground-truth binary activity indicator functions:
\begin{equation}
L_v(\bxi_2) = \frac{1}{N} \frac{1}{A}\sum_{n=1}^N \sum_{a=1}^{A}
\bigg( \frac{1}{t_{n,a}}\sum_{t=1}^{t_n} [\hat{m}_{a,t}=1] \cdot |v_{a,t} - \hat{v}_{a,t}| \bigg)
\label{eqn:viseme_loss}
\end{equation}
where $t_{n,a}$ is the number of frames where the rig parameter $a$ is active per clip $n$ in the ground-truth (i.e.,  $t_{n,a}=\sum_{t}[\hat{m}_{a,t}=1])$
An alternative approach would be to evaluate rig parameter differences when these are inactive (i.e., pushing the predictions towards $0$ in these cases). However, we found that this degrades the prediction quality of the rig parameters because the network over-focuses on making correct predictions in  periods where visemes are inactive. The smoothness loss $L_v'(\bxi_2)$ penalizes large   derivatives of predicted viseme and co-articulation parameters over time:
\begin{equation}
L'_v(\bxi_2) = \frac{1}{N} \frac{1}{A}\sum_{n=1}^N \sum_{a=1}^{A}\bigg( \frac{1}{t_{n,a}}\sum_{t=1}^{t_n} [\hat{m}_{a,t}=1] \cdot |\dot v_{a,t} | \bigg)
\label{eqn:viseme_smooth_loss}
\end{equation}
Finally, the loss function $L_j(\bxi_3)$ measures absolute differences between the training values  $\hat{\by}_{t}$ and predicted   values  ${\by}_{t}$ of the 2D JALI viseme field, while $L_j'(\bxi_3)$ penalizes large   changes in the viseme field values over time:
\begin{equation}
L_j(\bxi_3) = \frac{1}{N}\sum_{n=1}^N \bigg( \frac{1}{t_n}\sum_{t=1}^{t_n} ||\by_{t} - \hat{\by}_{t}||_1 \bigg)
\label{eqn:jali_loss}
\end{equation}
\begin{equation}
L_j'(\bxi_3) = \frac{1}{N}\sum_{n=1}^N \bigg( \frac{1}{t_n}\sum_{t=1}^{t_n} ||\dot \by_{t}||_{1} \bigg)
\label{eqn:jali_smooth_loss}
\end{equation}

Minimizing the above multi-task loss function is done through batch gradient descent with with batch size $256$, learning rate $0.00001$, momentum set to $0.9$, over $200$K iterations.

\paragraph{Thresholding activation probabilities and hold-out validation.} In the training stage, we also compute a threshold $thr_a$ for each rig parameter $a$ that determines when to activate it based on the  rig control activation probability produced by our network, i.e., check  $P(m_{a,t} > thr_a)$. One potential choice could be to simply set $thr_a=0.5$. \rev{Yet, we found that optimizing the threshold per rig  parameter through a dense grid search in a small hold-out validation dataset (10\% our training dataset clips are used for hold-out validation) and selecting the value  yielding the best precision and recall in rig activations in that dataset offered better performance. The same hold-out validation set and procedure are used to set the weights of the loss terms in Equations \ref{eqn:pretrain_loss} and \ref{eqn:mtl_loss}. }  

\paragraph{Implementation and running times.}
Our network is implemented in Tensorflow. Pre-training takes $30h$ and joint training takes $15h$ 
 in our training datasets measured on a TitanX GPU. At test time, audio feature extraction and network inference (forward propagation) is performed at $100$ FPS ($10$ms per frame) with a lag of $120$ ms relative to the current frame (see Section \ref{sec:technical},  phoneme branch paragraph).
\rev{Our code for training and testing the network, trained models, datasets, and results are available on our project web page.}~\footnote{\url{http://people.umass.edu/~yangzhou/visemenet}}

\vskip -2mm
\section{Evaluation}
\label{sec:results}
\label{sec:evaluation}

We evaluated our method and alternatives both quantitatively and qualitatively. In this section, we primarily focus on quantitative evaluation. We refer the reader to the video for qualitative results and comparisons. 
\paragraph{ Methodology.} We focused on the BIWI 3D\ audiovisual  dataset to validate our method and alternatives. As mentioned in the previous section, exemplar JALI-based motion curves were provided by an artist for the whole dataset, thus we can compare predicted rig parameters  to ground-truth. In addition, each of the $14$ speakers of the BIWI dataset speaks the same sentence in both neutral and  expressive styles, conveying emotions such as anger, sadness, fear, nervousness, and excitement. Thus, we can  compare our method and alternatives on how well they handle different styles of speech.  

Because this JALI-annotated dataset has only $14$ speakers, we perform the evaluation through a leave-one-out approach: for each of the $14$ BIWI speakers, we perform pre-training  on our audiovisual dataset (including GRID, SAVEE, and BIWI but excluding that BIWI\ speaker), and then  perform joint training on the JALI-annotated BIWI\ dataset using the other $13$ speakers. As a result, we form  $14$ training and test 
splits, where the test splits always involve a speaker not observed during training. Since we aim at learning a speaker-independent, generic model, we believe that this is a more compelling and practically useful generalization scenario, compared to training and testing on the same speaker.

 
\paragraph{Quantitative evaluation measures.} The first evaluation measure we use is the rig parameter  \emph{activation precision}, which measures how often we activate the right viseme and co-articulation rig parameters based on the ground-truth. Specifically, given a binary variable $\hat{m}_{a,t}$ indicating whether the rig parameter $a$ is activated or not at frame $t$ in the ground-truth, and given our predicted  binary variable $m_{a,t}$ for that parameter and frame, the precision is calculated as the number of times we correctly predict activations for the rig parameters (i.e., $\sum_{a,t}[\hat{m}_{a,t}=1 \,\&\, m_{a,t}=1]$, or in other words, the number of true positives) normalized by the total number of  predicted activations (i.e., $\sum_{a,t}[m_{a,t}=1]$). We also evaluate the rig parameter \emph{activation recall}, which measures out of all the ground-truth rig parameter activations, what fraction of them we predict correctly. The recall is calculated as the number of times we correctly predict activations for the rig parameters  (again, number of true positives) divided by the total number of ground-truth activations $\sum_{a,t}[\hat{m}_{a,t}=1]$. In the  ideal scenario,  precision and recall should be both $100\%$. 

We also measure the \emph{motion curve differences}, which evaluates the absolute differences of our predicted rig parameter values (viseme, co-articulation, and JALI field parameters) compared to their ground-truth values averaged over  the test frames where the corresponding rig parameters are active either in the ground-truth or in the predictions.  \rev{We note that we do not consider inactive parameters in the evaluation of motion curve differences because the motion curves are very sparse; zero-values would dominate the measure otherwise}.
Since all the rig parameters are normalized between $[0,1]$ during training and testing, these differences can  be treated as percentages. 


\begin{table}[t!]
\caption{\rev{Precision and recall for activation of rig controls for our full method and degraded versions of it for neutral and expressive speech styles, averaged over all test splits (higher precision and recall are better). We also report the average standard deviation (SD) of precision and recall for each variant.}}
\vskip -4mm
\centering
\begin{tabular}{@{}c@{}|@{}c@{}||@{}c@{}||@{}c@{}||@{}c@{}||@{}c@{}||@{}c@{}|}
                & \multicolumn{3}{c||}{neutral}  &  \multicolumn{3}{c|}{expressive } \\
                \cline{2-7}
                & \,precision\, & \,recall\, & \,SD\,  &  \,precision\,   & \,recall\, & \,SD\,  \\ 
                & (\%)            & (\%)    & \       & (\%)               & (\%)     & \       \\ 
\hline
 full method            & \textbf{89.5} & \textbf{92.2} & 1.9  & \textbf{90.1}  & \textbf{92.3} & 2.2 \\
landmark-based         & 73.6          & 82.0          & 5.1  & 74.4           & 82.7          & 4.9\\
phoneme-based          & 87.8          & 91.5          & 1.8  & 88.2           & 91.6          & 2.0\\  
audio-based             & 68.7          & 81.3          & 5.2  & 69.3           & 82.2          & 4.9\\ 
no transfer learning    & 85.8          & 89.5          & 2.3  & 86.1           & 89.6          & 2.2\\
no shared weights (LP)\, & 88.5          & 91.5          & 1.0  & 88.8           & 91.6          & 1.9\\ 
shared weights (V)\,      & 89.0          & 91.9          & 1.7  & 89.3           & 91.9          & 1.9\\ 
ASR-based\,             & 87.6          & 89.2          & 1.6  & 88.4           & 89.8          & 1.4\\ 
GRU-based\,                   & 87.4          & 91.2          & 2.1  & 87.8           & 91.1          & 2.3\\ 
sliding window-based\,  & 78.1          & 77.8          & 1.5  & 78.6           & 78.2          & 1.5\\ 
\hline
\end{tabular}
\label{tab:measures1}
\end{table}

\begin{table}[t!]
\caption{\rev{Percentage difference of motion curves (including viseme, co-articulation, and JALI field parameters) for our method and degraded versions of our architecture for neutral and expressive styles of speech, averaged over all test splits (lower difference is better). We also report the standard deviation (SD) of the percentage differences for each variant.}}
\vskip -4mm
\centering
\begin{tabular}{@{}c@{}|@{}c@{}||@{}c@{}||@{}c@{}||@{}c@{}|}
                &\multicolumn{2}{c||}{neutral}  &  \multicolumn{2}{c|}{expressive }  \\
                \cline{2-5}
                & motion curve    & \,\,SD\,\,          & motion curve  & \,\,SD\,\,  \\ 
                & \,\, differences (\%)\,\, & \ &\,\, differences (\%)\,\, & \  \\ 
\hline
 full method            & \textbf{7.8} & 0.8  & \textbf{7.6} & 0.9 \\
landmark-based         & 13.6         & 1.8  &     13.2     & 1.5    \\
phoneme-based          & 9.0          & 0.7  &     8.8      & 0.7   \\  
audio-based             & 14.5         & 1.8  &     14.2     & 1.6    \\ 
no transfer learning    & 9.7          & 0.9  &     9.6      & 0.8   \\
no shared weights (LP)    & 8.8          & 0.6  &     8.7      & 0.7\\ 
shared weights (V)        & 9.5          & 0.7  &     9.2      & 0.7\\ 
ASR-based\,             & 9.1          & 0.6  &     8.8      & 0.6     \\ 
GRU-based\,                   & 9.1          & 0.7  &     9.0      & 0.8     \\ 
sliding window-based    & 15.4         & 0.3  &     15.2     & 0.4\\ 
\hline
\end{tabular}
\label{tab:measures2}
\end{table}

\paragraph{Quantitative Comparisons.} We compare our network with the following alternative architectures: (a) \textbf{landmark-based:} we eliminate the phoneme group stage of our network  i.e., visemes are predicted based on  landmarks and audio features only, (b) \textbf{phoneme-based:} we eliminate the landmark stage of our network i.e.,  visemes are predicted based on phoneme groups and audio features only, (c) \textbf{audio-based:} we eliminate both the landmark and phoneme group stages i.e., visemes are predicted directly from audio features alone, which also implies that there is no pre-training since no training phoneme groups or landmarks  can be used in this condition (d) \textbf{no transfer learning:} we keep all the stages of our network, yet we train only on the JALI-annotated BIWI training split and not on the rest of the audiovidual datasets, \rev{(e) \textbf{no shared weights (LP):} we disable weight sharing  between the landmark and phoneme stages i.e., the LSTM layers of these two stages have independent parameters in this variant, (f) \textbf{shared weights (V):} we force weight sharing between the layers of the three LSTM modules used for predicting rig control activations, viseme/co-articulation parameters, and JALI parameters in the viseme stage (this is in contrast to our proposed architecture that uses  LSTMs without shared weights in this stage).} (g) \textbf{ASR-based:} instead of using our phoneme group prediction part, we pass the input audio through the popular Google Cloud automatic speech recognition engine \cite{googlecloud} to extract text, then extract phonemes based on the MFA forced aligner \cite{mcauliffe2017montreal}, form the phoneme groups of Figure \ref{fig:viseme_list}, encode them into a binary vector with $1$s corresponding to present phoneme groups and $0s$ for the non-present ones  per frame, and pass this vector to our viseme branch (instead of our phoneme group representations). This alternative architecture tests the condition where phonemes are acquired automatically in separate stages through existing off-the-shelf tools, \rev{(h) \textbf{GRU-based:} we use Gated Recurrent Units (GRUs) \cite{Cho2014} instead of LSTMs as  memory modules,} (i) \rev{\textbf{sliding window-based:} instead of using LSTMs in our three stages, we experimented with the memory-less neural network modules based on three fully connected hidden layers operating on sliding windows, as proposed in  \cite{Taylor:2017:DLA}. We note that we further benefited this approach by using the same type of inputs in each stage (i.e., the viseme stage receives audio, landmarks, and phonemes as input instead of using phonemes alone as done in  \cite{Taylor:2017:DLA}) and also using the same pre-training and transfer learning procedure as in our approach (without  these enhancements, the performance was  worse). We also increased the number of hidden nodes per layer  so that the number of learnable parameters is comparable to the one in our architecture (using the original number of hidden nodes also resulted in  worse performance).  } 

We trained these  alternative architectures based on the same corresponding loss functions in the same training sets as our method,   performed the same hyper-parameter tuning procedure as in our method, and evaluated them in the same test splits. Table \ref{tab:measures1} and Table \ref{tab:measures2} report the abovementioned evaluation measures for our method and the alternatives for  neutral and expressive styles of speech. Our full method offers the best performance in terms of all evaluation measures and different styles of speech. Based on the results, if one attempts to skip our intermediate phoneme group and landmark stages, and predict visemes from audio features directly (``audio-based'' condition), then the performance degrades a lot (see also accompanying video and Figure \ref{fig:alternative_results_failures} for qualitative comparisons). Skipping the phoneme group stage (``landmark-based'' condition) also results in a large performance drop, which indicates that recognizing phoneme groups is crucial for predicting correct visemes, as the psycho-linguistic literature indicates. Skipping landmarks (``phoneme-based'' condition) also results in a noticeable drop in performance for both neutral and expressive styles of speech. Using off-the-shelf tools for viseme recognition (``ASR-based'' condition) also results in worse performance than our approach. Note also that this ASR-based approach is language-dependent and requires an input language-based phoneme model specification, while our approach is language-agnostic since our phoneme groups are based on the International Phonetic Alphabet (IPA). \rev{Furthermore, we observed that transfer learning and weight sharing in the landmark and phoneme stages improve performance}. \rev{Using GRUs results in  worse performance compared to LSTMs.}
\rev{Finally, replacing the LSTMs  with  fully connected network modules operating on sliding windows  causes a large drop in performance.}   

\paragraph{Qualitative comparisons.}  Our video shows facial animation results produced by our method and degraded versions of our architecture as well as comparisons with  previous works  \cite{Karras:2017:AFA}, \cite{Suwajanakorn:2017:SOL}, and \cite{Taylor:2017:DLA}. Quantitative comparisons with these previous works are not possible because their used test rigs are not FACS-enabled, and their implementation is not publicly available. In contrast to our approach,  none of these previous methods produce editable, animator-centric viseme curves or facial action units.  \rev{We also demonstrate generalization to speech animation involving different languages.}

\vskip -2mm
\section{Conclusion}

We presented an animator-centric, deep learning approach that maps audio to speech motion curves. 
There are various avenues for future work. Our implementation currently uses hand-engineered audio features. Replacing them with learned features, similarly to what is done in image and shape processing pipelines, could help improving performance. Another interesting extension would be to incorporate a discriminator network that would attempt to infer the quality of the generated animations, and use its predictions to boost the performance of our viseme generator network, as done in cGAN-based approaches \cite{pix2pix2016} for image and shape synthesis. Finally, our method is able to drive only the lower part of the face. Learning to control the upper face e.g., eyes, without explicit supervisory signals would also be a fruitful direction.

The marriage between animator-centric techniques and deep-learning has the potential to fundamentally alter current facial animation practice in film and game studios, leaving animators free to focus on the creative and nuanced aspects of character expression. We believe our solution is a significant step in this direction.

\vskip -2mm
\section*{Acknowledgements}
We acknowledge support from  NSERC and NSF  (CHS-1422441, CHS-1617333, IIS-1617917). 
We thank Pif Edwards and anonymous reviewers for their valuable feedback. Our  experiments  were  performed in the UMass GPU cluster obtained under  the  Collaborative  R\&D  Fund  managed  by  the  Massachusetts Technology Collaborative.

\bibliographystyle{ACM-Reference-Format}
\bibliography{audio2visemes}

\end{document}